\documentclass[fleqn,usenatbib]{mnras}

\usepackage[T1]{fontenc}
\DeclareRobustCommand{\VAN}[3]{#2}
\let\VANthebibliography\thebibliography
\def\thebibliography{\DeclareRobustCommand{\VAN}[3]{##3}\VANthebibliography}

\usepackage{graphicx}	% Including figure files
\usepackage{amsmath}	% Advanced maths commands
\usepackage{amssymb}	% Extra maths symbols

\title[The radio signal detected over the whole pulse phase]{The radio signal of PSR B0950$+$08 is detected over the whole pulse phase}

\author[Zhengli Wang et al.]{
Zhengli Wang,$^{1}$
Jiguang Lu$^{2,3}$\thanks{Email: lujig@naoc.cas.cn},
Jinchen Jiang,$^{4}$
Jie Lin,$^{4}$
Kejia Lee,$^{4,5}$
Enwei Liang$^{1}$
and Renxin Xu$^{4,5}$\thanks{Email: r.x.xu@pku.edu.cn }
\\
$^{1}$Guangxi Key Laboratory for Relativistic Astrophysics, School of Physical Science and Technology,
Guangxi University, Nanning 530004, China\\
$^{2}$National Astronomical Observatories,
Chinese Academy of Sciences, Beijing 100012, China\\
$^{3}$Guizhou Radio Astronomical Observatory, Guizhou 550025, China\\
$^{4}$Department of Astronomy, School of Physics,
Peking University, Beijing 100871, China\\
$^{5}$Kavli Institute for Astronomy and
Astrophysics, Peking University, Beijing 100871, China
}

\date{Accepted XXX. Received YYY; in original form ZZZ}

\pubyear{2021}

\begin{document}
\label{firstpage}
\pagerange{\pageref{firstpage}--\pageref{lastpage}}
\maketitle

%\textbf{The radio signal of PSR B0950$+$08 is detected over the whole pulse phase}

\begin{abstract}
Pulsars are the ``lighthouses'' in the universe. Periodic pulses with the duty-cycle $\sim 10\%$ are detected when the radio beam of the rotating pulsar sweeps across the telescope.
%
%are thought to emit periodic pulses with duty-cycle $\sim 10\%$.
%
In this report, the 160\,min-data of a nearby pulsar, PSR B0950$+$08, observed with the Five-hundred-meter Aperture Spherical radio Telescope (FAST) is analysed.
Thanks to the extremely high sensitivity of FAST, it is found that the radiation of PSR B0950$+$08 could be detected over the entire pulse period.
To investigate the radiative characteristics of the pulsar's ``bridge emission'', a function, $\Theta(n)$, is defined to reveal the weak radiation there.
It is suggested that the narrow peaks of both the main and the inter pulses could be radiated at low altitude, while other weak emission (e.g., the ``bridges'')
%from high magnetosphere far away from the surface though its radiative mechanism is still a matter of debate.
from upper magnetosphere though its radiative mechanism is still a matter of debate.
The measured mean pulse behaviors are consistent with previous results in
the phase of strong emission of this pulsar, 
and both the frequency-independent separation between the inter-pulse and main pulse and the narrow pulse width may support a double-pole model.
%
%Nonetheless, in order to finalize  the magnetospheric geometry, further polarization observation with FAST is surely required.
%which would only be believable in the phase of weak emission if the baseline is determined with certainty in the future.
%\textbf{which would be well disentangled the magnetospheric geometry of this pulsar if the polarization characteristics of the weak emission phase are detected via the determination of the baseline.}
In order to understand the magnetospheric geometry of this pulsar, further polarization calibrated observation with FAST and a proper determination of the baseline emission, especially during the weak emission phase, are surely required.
\end{abstract}


%%%%%%%%%%%%%%%%%%%%%%%%%%%%%%%%%%%%%%%%%%%%%%%%%%

%%%%%%%%%%%%%%%%% BODY OF PAPER %%%%%%%%%%%%%%%%%%

\section{Introduction}

The nearby, bright pulsar, PSR B0950$+$08, is well known, with a spin period of 253 ms and dispersion-measure (DM) of 2.97 pc/cm$^3$.
Abundant observations have already been obtained for this normal pulsar~\citep[e.g.][]{1991ApJ...373L..17H,1998A&AS..127..355K,2001ApJ...553..341E,2004A&A...418..203S,2005MNRAS.364.1397J}.
For instance, the averaged pulse profile of PSR B0950$+$08 shows frequency-dependent properties~\citep[e.g.][]{1991ApJ...373L..17H,1998A&AS..127..355K}, and~\citet{2004A&A...418..203S} reported that the frequency dependent sinusoidal modulation of the pulse shape may result from the Faraday rotation effect.
Also, its linear polarization is shown in the literature~\citep[e.g.][]{2001ApJ...553..341E,2004A&A...418..203S,2005MNRAS.364.1397J}, but is too low to fit the polarization position angle (PPA) at the main pulse longitude by typical 'S' shape in the Rotating Vector Model~\citep[RVM,][]{1969ApL.....3..225R}.
In fact, ~\citet[][]{2001ApJ...553..341E} presented the RVM fitting for PSR B0950$+$08. When they fitted the PPA of the whole pulse longitudes (Fig.8), the longitude range (-10,+15)$^{\circ}$ of distortion regions appearing at the main pulse must be unweighted.
~\citet{2002A&A...396..171P} confirmed the characteristic timescales of micro-structure and narrow micro-structure are orders of 0.1\,ms and 10\,$\mathrm{\mu}$s, respectively, while the timescale of fine structure is $\sim 1$ ms~\citep{2016MNRAS.455..150U}.
 Besides, the separation between inter-pulse and main pulse is frequency independent below 5 GHz, and this feature demonstrates that multi-pole magnetic field may contribute significantly in order to understand the ``bridge'' emission~\citep{1981ApJ...249..241H,2002AAS...20111802N}.
 %
 %\textbf{Also, the averaged pulse profile shows complex shapes, and its radio emission beam is described by the so-called ``core-cone'' model. According to this model, the beam consists of a central core surrounded by two rings of nested conal emission, which are named the inner and outer cones, respectively \citep{1983ApJ...274..333R, 1990ApJ...352..247R}.}

In addition to the mean pulse profile, the single pulse behaviors have been studied for PSR B0950$+$08.
%EDF Statistics for Goodness of Fit and Some Comparisons
After a detailed study of intensity dependence of individual pulses, ~\citet{2002AAS...20111802N} found a possible correlation between the intensities and locations of the inter-pulse with the leading component of the main pulse.
Additionally, PSR B0950$+$08 also exhibits the nulling pulse phenomenon and the sharp flux variation in a short-time scale~\citep[e.g.][]{2004ApJ...610..948C,2004A&A...418..203S,2012AJ....144..155S}. Besides, the giant pulse and giant micro-pulse phenomena have also been detected from PSR B0950$+$08~\citep[e.g.][]{2004ApJ...610..948C,2012AJ....144..155S,2015AJ....149...65T}, but the origin of them is still a mystery.

Nevertheless, both properties of the mean pulse and single pulse would be significantly influenced by the baseline determination.
%as the conventional way of baseline subtraction (i.e., subtracting the intensity of the minimum level of pulse phase) can hardly be acceptable for a pulsar with unusually wide pulse like that of PSR B0950$+$08.
The conventional baseline subtraction is to subtract the weak emission region away from the strong emission phase (i.e., the main pulse) of pulsar.
If this baseline subtraction is directly used to subtract the weak emission region of the pulsar with unusually wider pulse such as PSR B0950$+$08, the impulsive radio signal of the weak emission phase of this pulsar is also subtracted.
More details on the discussion of the baseline will also be presented in the latter section.
%, and this method is usually adopted for the narrow pulse phase radiative pulsar. There is a flat platform at the low-level emission region of these pulsars, and the radio signal of the flat platform is only contributed by the receiver system and the pulsar wind nebulae (PWN).}
%
%Therefore, if this baseline subtraction method is used to subtract the baseline region for a pulsar with unusually wide pulse like that of PSR B0950$+$08, the impulsive radio signal of the low-level emission region such as
%
%the ``bridge'' region of this pulsar is also to be subtracted.}
% Because there is a clear concave structure at the region away from the main pulse for PSR B0950$+$08.}
%Because the conventional way of baseline subtraction is to subtract the intensity of the minimum level of pulse phase.}
%
~\citet{1981ApJ...249..241H} strongly supported the conclusion that the radio emission from PSR B0950$+$08 occurs over at least $ 83\%$ of the rotation period.
They pointed that, because of limited sensitivity of telescope, the radio emission characters of the ``two minimum levels'' at the bridge regions between inter-pulse and main pulse (the remaining $17 \%$ pulse longitude) could not be defined with certainty.
%\textbf{Therefore, if the conventional baseline subtraction is directly used to subtract the low-level emission of the pulsar with unusually wider pulse such as PSR B0950$+$08, the impulsive radio signal of the weak emission phase of this pulsar is also subtracted.
%
%Besides, the baseline subtraction is very import to obtain the PPA, and the correct PPA is very useful for understanding the magnetospheric geometry of the pulsar.
%
%If the emission region (with the pulsed radio signal) is regarded as the baseline region and then subtract it.
%
%The polarization characteristics of this region, and even the whole pulse phase may be also influenced by the baseline subtraction. More details discussing the baseline and the PPA will also be provided in the later discussion section.}
%

Up to now, similar properties have also been discovered in PSR B1929$+$10 and some other millisecond pulsars~\citep[e.g.][]{2004MNRAS.351..808M,2015MNRAS.449.3223D},
%and all these pulsars have been reported that the emission
for all these pulsars it has been reported that the emission covers an unusually wide range of pulse longitude.
Therefore, the radio emission of the whole pulse phase for these pulsars is worth detecting.
~\citet[][]{1995ApJ...455L..55N} discovered a binary millisecond pulsar, PSR J0218$+$ 4232, with the extremely broad pulse profile during imaging observations, they later confirmed that this pulsar exhibits a significant fraction of the radio emission is not pulsed.
Meanwhile, ~\citet[][]{2020MNRAS.495.2125R} reported the detection result of the emission features of off-pulse emission (the emission is non-periodic and stationary) from PSR B0950$+$08. Through imaging analysis, they concluded that the off-pulse emission of this pulsar is mainly originated from the outside of the magnetosphere.
Nevertheless, it is hard to completely rule out the possibility of magnetospheric origin for off-pulse emission via imaging.
The difference between the emission origins from inside and outside of the magnetosphere is whether the radiation is impulsive and periodic, and in this work, the impulsive characteristic of PSR B0950$+$08 is attempted to be detected.
However, not all radio pulsars have off-pulse emission, ~\citet{2019A&A...627L...2M} presented a detailed detection of off-pulse emission for B0525$+$21 and B2045$-$16.
%
%Through analyzing the data of the European Very Long Baseline Interferometry Network (EVN),
They reported that the off-pulse emission above three times the rms noise levels for B0525$+$21 and B2045$-$16 are not detected using the European VLBI Network.
%
%they pointed that off-pulse emission of B0525$+$21 and B2045$-$16 (more than three times the rms noise levels) were not detected.
%this detection method
%
%Meanwhile, ~\citet[][]{2020MNRAS.495.2125R} reported the detection result of the emission features of off-pulse emission from PSR B0950$+$08. Through imaging analysis, they concluded that the off-pulse emission of this pulsar is mainly originated from the outside of the magnetosphere.
%
The imaging method is potentially used to determine the pulsars with wide pulse profile such as PSR B1929$+$10 in the future.
%
%In addition, it is hard to completely rule out the possibility of magnetospheric origin for off-pulse emission via imaging.
%
%The difference between the emission origins from inside and outside of the magnetosphere is whether the radiation is impulsive and periodic, and in this work, the impulsive characteristic of PSR B0950$+$08 is attempted to be detected.}
%
%\textbf{The imaging method is potentially used to determine the pulsars with wide pulse profile such as PSR B1929$+$10 in the future.}
%

\begin{figure*}
	\centering
	\includegraphics[width = \textwidth]{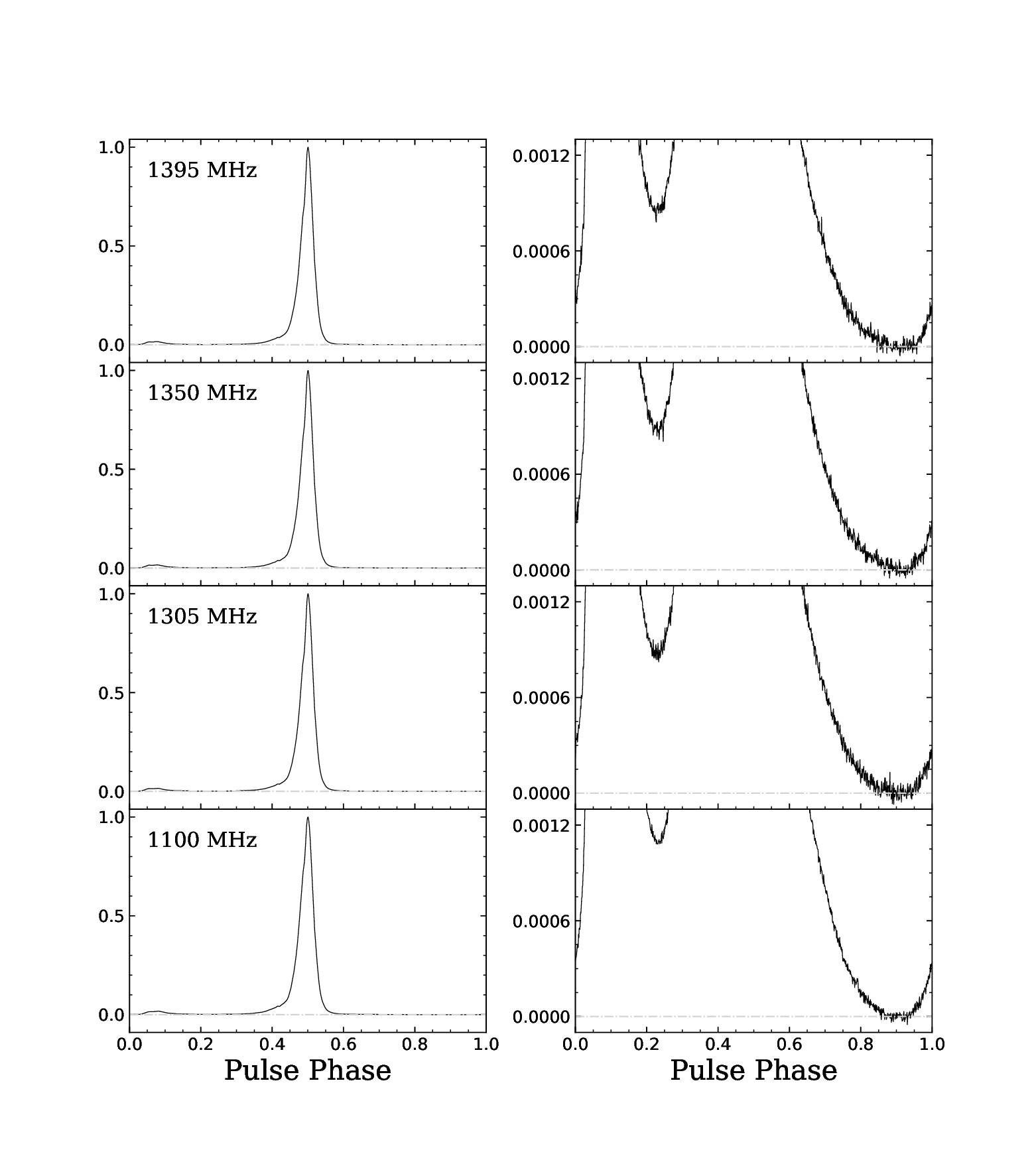}
    \caption{The averaged pulse profile at 1100, 1305, 1350, and 1395\,MHz are shown in the left panel, and the corresponding $\times$800 expanded scale views are plotted in the right panels. The emission region of the pulse phase from 0.88 to 0.92 is regarded as the baseline region.}
    \label{fig1}
\end{figure*}

The largest
single-dish radio telescope in the world,
%
%China's five-hundred-meter aperture
%spherical radio telescope (FAST),
%
china's FAST would
be an appropriate tool to understand more about radio emission as well as the radiative mechanism of PSR B0950$+$08,
taking advantage of its extremely high sensitivity~\citep{2019SCPMA..6259502J,2020RAA....20...64J}.
New scientific achievements have already been obtained with FAST, including studies of pulsars~\citep[e.g.][]{LuJG2019} and fast radio bursts~\citep[e.g.][]{XuR2021}.
In this article, the emission from the ``bridge'' of PSR B0950$+$08 is determined (as shown in Fig.~\ref{fig1}), exhibiting a concave structure.
In this paper, a detailed study of the radiation properties of the entire $360^{\circ}$-pulse longitude is presented. %
The observations with FAST and the data reduction are described in Section~\ref{obs}.
To determine the emission at the entire pulse period, a depth data analysis and the radiation characteristics of this pulsar are presented in Section~\ref{dat}.
Finally, general conclusion and discussions are provided in Section~\ref{con}.

%\begin{table}
%	\centering
%	\caption{This is an example table. Captions appear above each table.
%	Remember to define the quantities, symbols and units used.}
%	\label{tab:example_table}
%	\begin{tabular}{lccr} % four columns, alignment for each
%		\hline
%		A & B & C & D\\
%		\hline
%		1 & 2 & 3 & 4\\
%		2 & 4 & 6 & 8\\
%		3 & 5 & 7 & 9\\
%		\hline
%	\end{tabular}
%\end{table}

\section{OBSERVATION AND DATA REDUCTION}
\label{obs}

FAST  is located in Guizhou, a southwestern province of China, at longitude $106.9^{\circ}$ E and latitude $25.7^{\circ}$ N.
In this work, the 19-beam receiver system was adopted.
The central frequency and bandwidth of the receiver are 1250 and 400\,MHz, respectively~\citep{2019SCPMA..6259502J,2020RAA....20...64J}.
The system temperature of the 19-beam receiver is less that 24\,K for central beam, and the system stability for most observation modes with the observation zenith angle less than $26.4^{\circ}$, is $\sim 1 \%$ over 3.5\, hours~\citep{2019SCPMA..6259502J,2020RAA....20...64J}

In this work, PSR B0950$+$08 was observed with tracking mode on MJD 59396 (July 1st, 2021).
The data were recorded as the 8 bit-sampled search mode PSRFITS format~\citep{2004PASA...21..302H} with 4096 frequency channels, and the frequency resolution is $\sim$ 0.122MHz.
The entire integration time is 160 minutes, and the time resolution is $\sim$ 50\,$\mathrm{\mu}$s.
The DSPSR software package~\citep{2011PASA...28....1V} was adopted in the data reduction, and the individual pulses were generated with 1024 phase bins across the pulse period of 253\,ms~\citep{2004MNRAS.353.1311H}.
%
%\textbf{
%To eliminate the effect of the bandpass, the upper and lower edges of the bandpass are directly subtracted, where the edges account for $\sim 10 \%$ of the bandwidth.
%
%The data were processed to remove dispersion delay caused by the interstellar medium and generate single pulse archives with the DSPSR software packages~\citep[][]{2011PASA...28....1V}. The ephemeris of the pulsar was provided by the ATNF pulsar catalogue (PSRCAT, version 1.56)~\citep[][]{2005AJ....129.1993M}. And each single pulse includes 1024 phase bins across the pulse period of 253\,ms~\citep[][]{2004MNRAS.353.1311H}.}
 %

To ensure that the radio signal of the weak emission region of PSR B0950$+$08 is not affected by the radio frequency interference (RFI),
%and to quantitatively determine it, the RFIs are considered with
the RFI is detected using the dynamic spectrum. Significant and possible impulsive RFIs are eliminated.
Moreover, the narrow-band RFI is also efficiently mitigated by using the dynamics spectrum.
%
%Therefore, after eliminated RFI, the frequency channels of RFIs are eliminated, only these integrated sub-bands data with frequencies of 1100,
%
%The frequency channels of RFIs are eliminated, and
Then only the integrated sub-band data with frequencies of 1100, 1305, 1350, and 1395\,MHz is remained, and their bandwidth are 80, 50, 40, and 50\,MHz, respectively.
%
%The detailed measurement for the bandpass of the 19-beam receiver has been published in~\citep{2020RAA....20...64J}, and the bandpass spectrum has a mean amplitude around 22\,K. After the RFIs are efficiently mitigated by using the dynamics spectrum, all the remaining frequency channels only have little amplitude fluctuation, this effect can not affect the impulsive radio signal of the sub-pulse of PSR B0950$+$08.

In this work, in order to obtain the averaged pulse profile and to discuss the radiation characteristics of strong emission phase of PSR B0950$+$08, the emission region of the pulse phase from 0.88 to 0.92 (see Figs.~\ref{fig1} and~\ref{ppa}) is regarded as the baseline region and then subtract it.
Furthermore, to detect the pulsed radio signal of the ``bridge'' emission region of this pulsar.
The baseline in each single pulse can be determined according to the pulsed radio emission in each individual pulse is narrow pulse.
After the baseline subtraction i.e., subtracting the receiver contribution is carried out in each single pulse, and the radio emission caused by the systematic contribution is unpulsed. The systematic contribution of the ``bridge'' emission region is removed by the impulsive property of the sub-pulse of this pulsar.
More detailed detection of the pulsed radio signal across the whole pulse phase will be presented in the Section~\ref{dat} and ~\ref{con}.
%In order to ensure that the radio signal of the weak emission region of PSR B0950+08 is not the contribution of the receiver system and PWN.
%
%The non-linear evolution of the baseline with time needs to be determined.
%
%Considering that the radio signal of single pulse is detected over narrow pulse phase, the %baseline evolution with time can be eliminated through the flattening of the baseline evolution in %each individual.
%
%In this work, the baseline evolution with time is determined by averaging the 10 low-level %emission pulse phase bins in each single pulse. Additional discussions on this pulsar baseline are %also presented in Section~\ref{dat} and~\ref{con}}

%%%%%%%%%%%%%%%%%%%%%%%%%%%%%%%%%%%%%%%%%%%%%%%%%%%%%%%%%%%%%%%%%%%%%%%
\section{DATA ANALYSIS AND RADIATION CHARACTERISTICS}
\label{dat}

In order to discuss the emission characteristics of PSR B0950+08,
%the range of $\sim 4 \%$ of pulse phase (from 0.88 to 0.92 in Figs.~\ref{fig1} and~\ref{ppa}) is regarded as the region of minimum intensity. And
the emission from the 0.17 to 0.30 and 0.65 to 0.95 pulse phase intervals in Fig.~\ref{fig7}, which is called the low-emission regions in the texts, is defined as the bridge emission.
With this definition,
%it is  attempted to prove
we demonstrate that there is no off-pulse region in the pulse phase of PSR B0950$+$08 radio in 1050-1450\,MHz, and the emission, even though in the bridge region, is also impulsive.

%{This pulsed radiation in bridge region must be an overlapped pulse and anti-pulse.}
%of the bridge regions (from 0.65 to 0.95 pulse phase in Figure.~\ref{fig7}) are named the bridge emission.}
%
%Moreover, to eliminate the radio signals of the weak emission regions are not affected by the radio frequency interference (RFI) and to quantitatively determine it, the RFIs are considered with dynamic spectrum, significant and possible RFIs are identified and eliminated. The narrow-band RFIs had also been eliminated via dynamic spectrum.}
%
%\textbf{Therefore, after eliminated RFI, the frequency channels of RFIs are almost eliminated,
%
%only these integrated sub-bands data with frequencies of 1100, 1305, 1350, and 1395\,MHz are remained, and their bandwidth are respectively 80, 50, 40, and 50\,MHz.}
%

\begin{figure}
    \centering
    \includegraphics[width = \columnwidth]{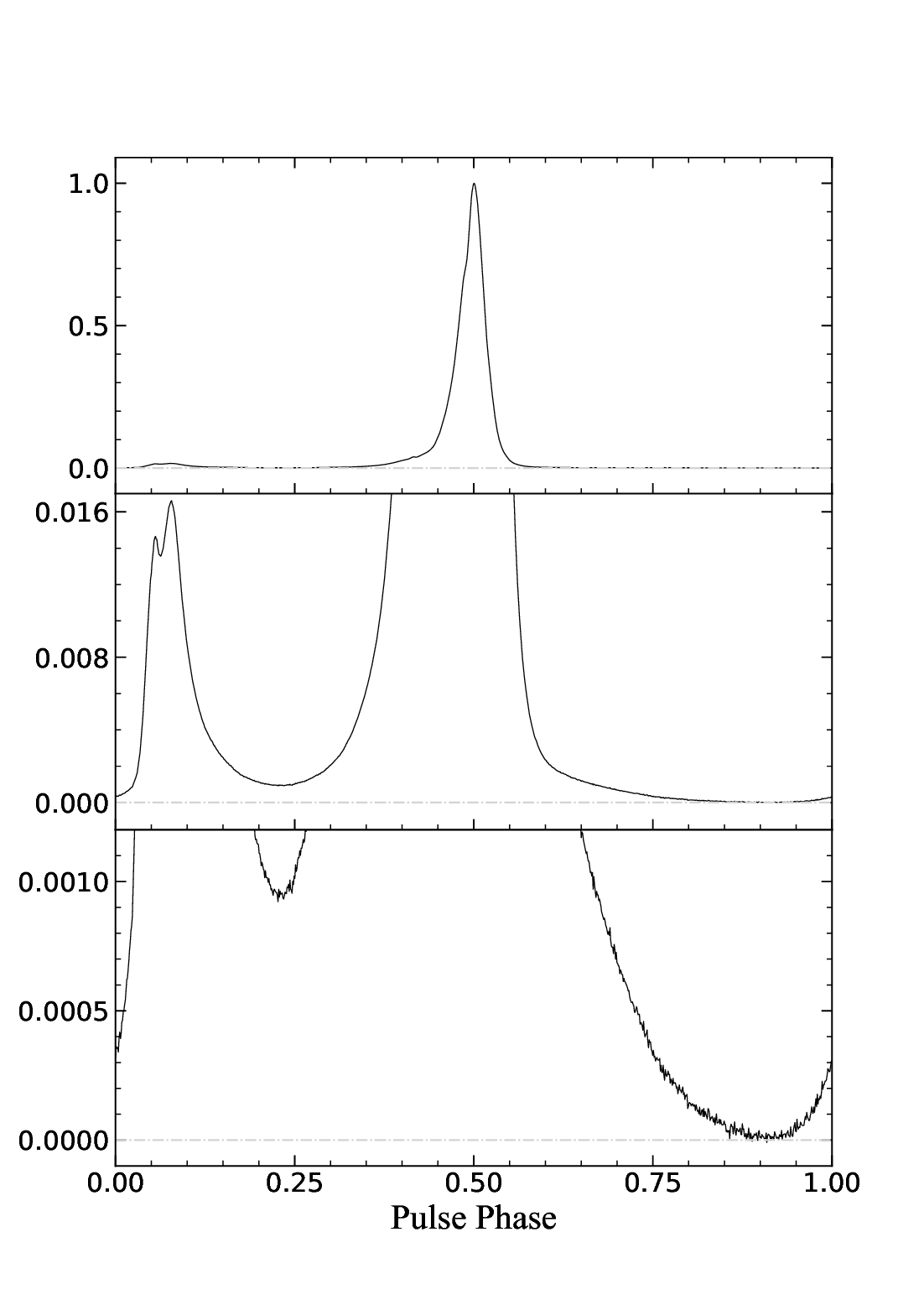}
    %\plotone{p.eps}
    \caption{The averaged pulse profile of the whole bandwidth is shown in the top panel, and the corresponding $\times$60 and $\times$900 expanded scale views are plotted in the middle and bottom panels, respectively. The emission region of the pulse phase from 0.88 to 0.92 is regarded as the baseline region.}
    \label{ppa}
\end{figure}

\begin{figure}
    \centering
    \includegraphics[width = \columnwidth ]{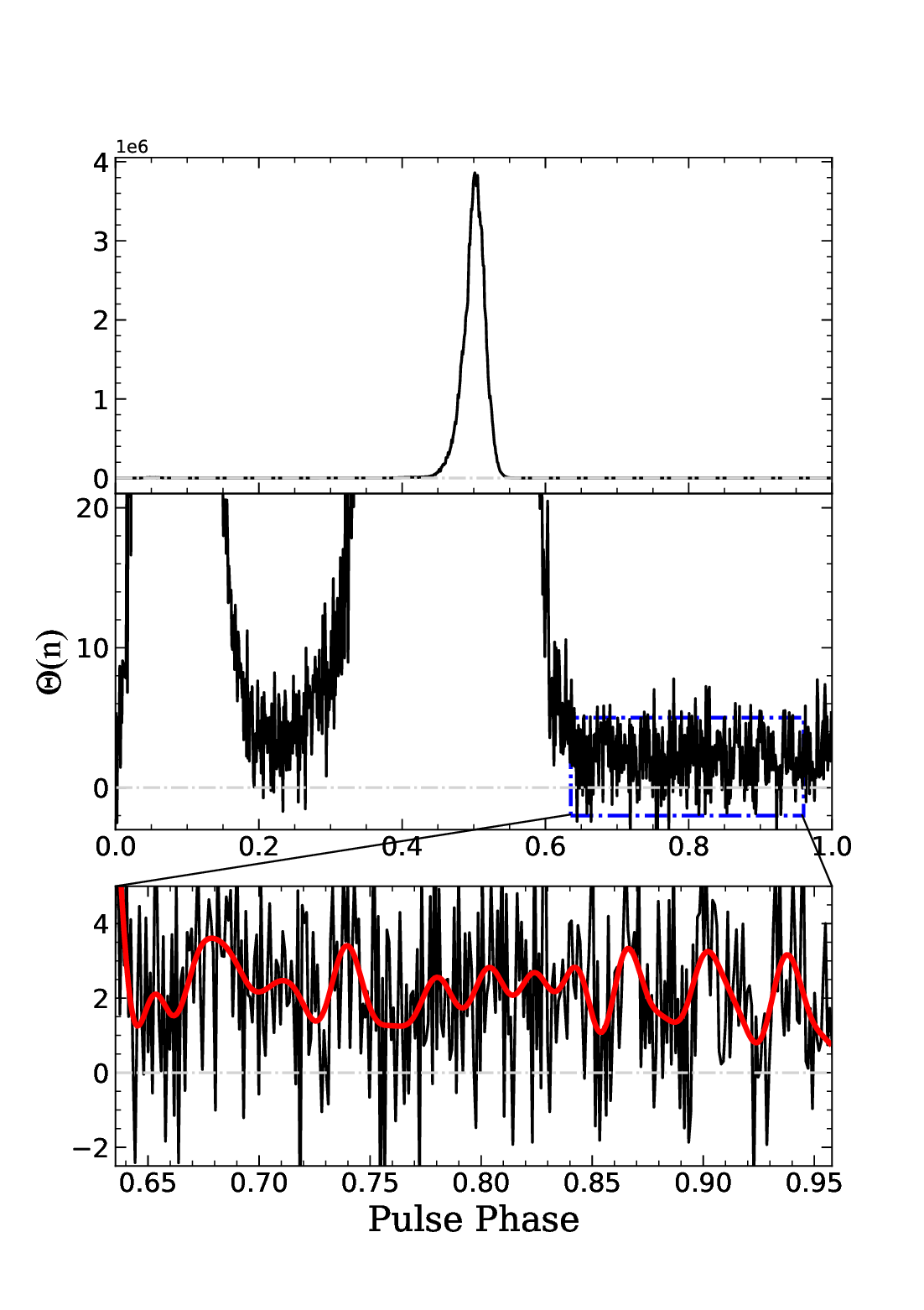}
    \caption{The $\Theta(n)$-function defined in equation (\ref{eq1}) of the whole pulse longitude are shown in the top panel, and vertically zoomed-in view is plotted in the middle panel. Moreover, the detailed property of the blue region is zoomed in the bottom panel.}
    \label{fig7}
\end{figure}

The averaged pulse profiles at all remaining frequencies are shown in Fig.~\ref{fig1}. As shown in Fig.~\ref{fig1}, the radio emission features across the entire pulse longitude are exhibited, and the observational properties of the two low-emission regions
%(the bridge components)
are clearly revealed from zoomed-in view in the right panels. The averaged pulse profile of the full bandwidth is also shown in Fig.~\ref{ppa}. Apparently, from the figure, there is a concave emission structure rather than flat platform at the two low-emission regions, the first concave structure appears at pulse phase of 0.25, meanwhile, another concave structure is revealed at pulse phase of 0.90.
This feature demonstrates a radio emission behaviour without off-pulse longitude, i.e., the pulsar radiates over the entire pulse period.

To reveal the emission characteristics of the two low-emission regions, a mathematical method is proposed to analyse the emission characteristics of the entire pulse longitude. The $\Theta(n)$-function is defined as following,
\begin{equation}
\begin{split}
    &\Theta(n) = \sum_{k =1}^{N_{\mathrm{period}}}\bigg\{  2 \times  \left( I_{ k,n-1} - I_{ k,n+1} \right)^2  \\
    & - \left[ \left( I_{ k,n} - I_{ k,n-1} \right)^2 + \left( I_{ k,n} - I_{ k,n+1}\right)^2 \right] \bigg\},
\end{split}
\label{eq1}
\end{equation}
where $I_{k,n}$ represents the signal intensity contributed by the $k$-th pulse and the $n$-th pulse phase bin, and $N_{\mathrm{period}} = 36746 $ is the pulse numbers. The results are shown as Fig.~\ref{fig7}.

\begin{figure}
    \centering
    \includegraphics[width = \columnwidth ]{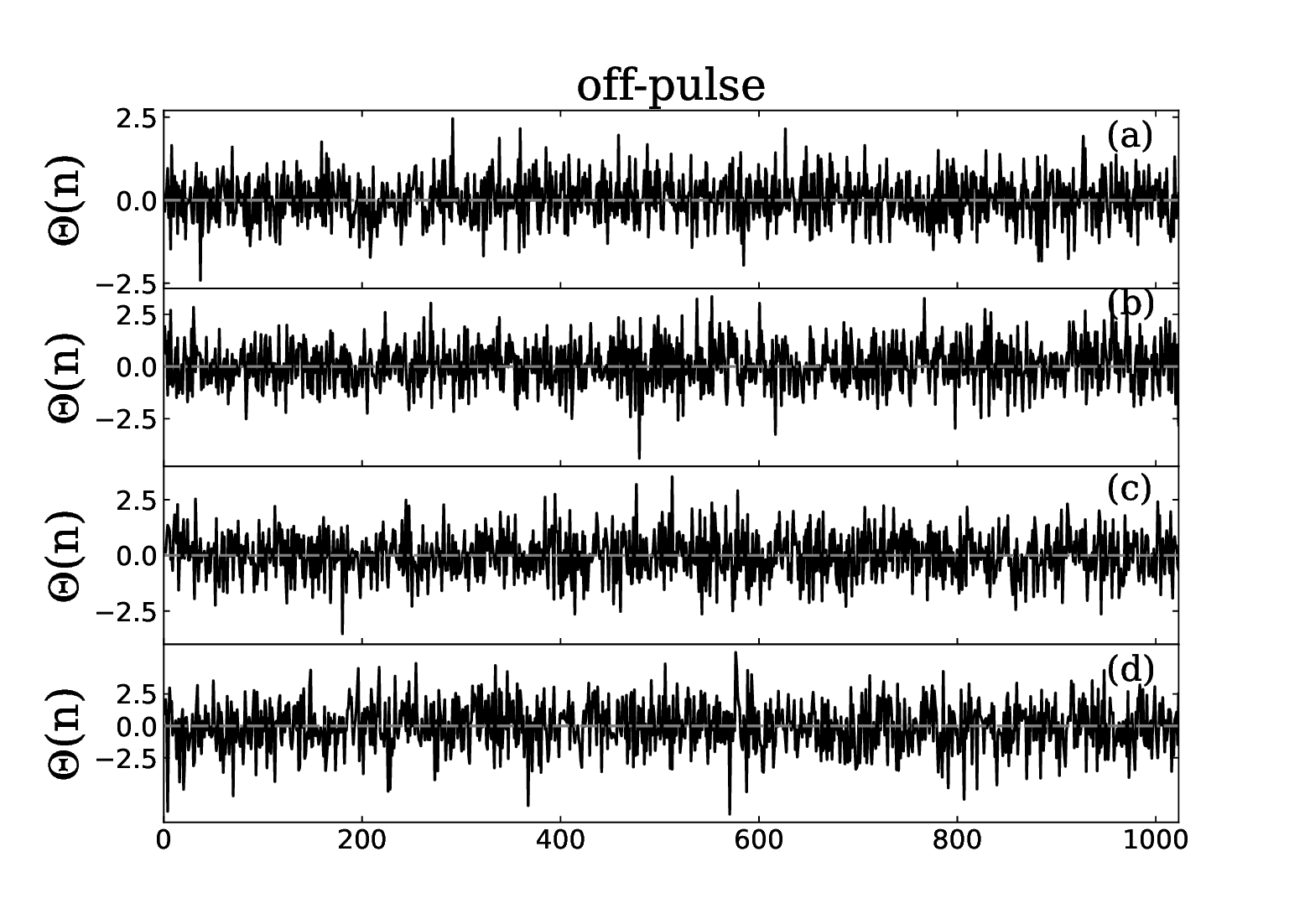}
    \caption{The behaviour of $\Theta(n) $-function for the simulated random data. Panels (a) and (b) correspond to gamma distributions with different shape parameters, and panels (c) and (d) represent log-normal and Gaussian distributions, respectively.}
    \label{off}
\end{figure}

\begin{figure}
    \centering
    \includegraphics[width = \columnwidth ]{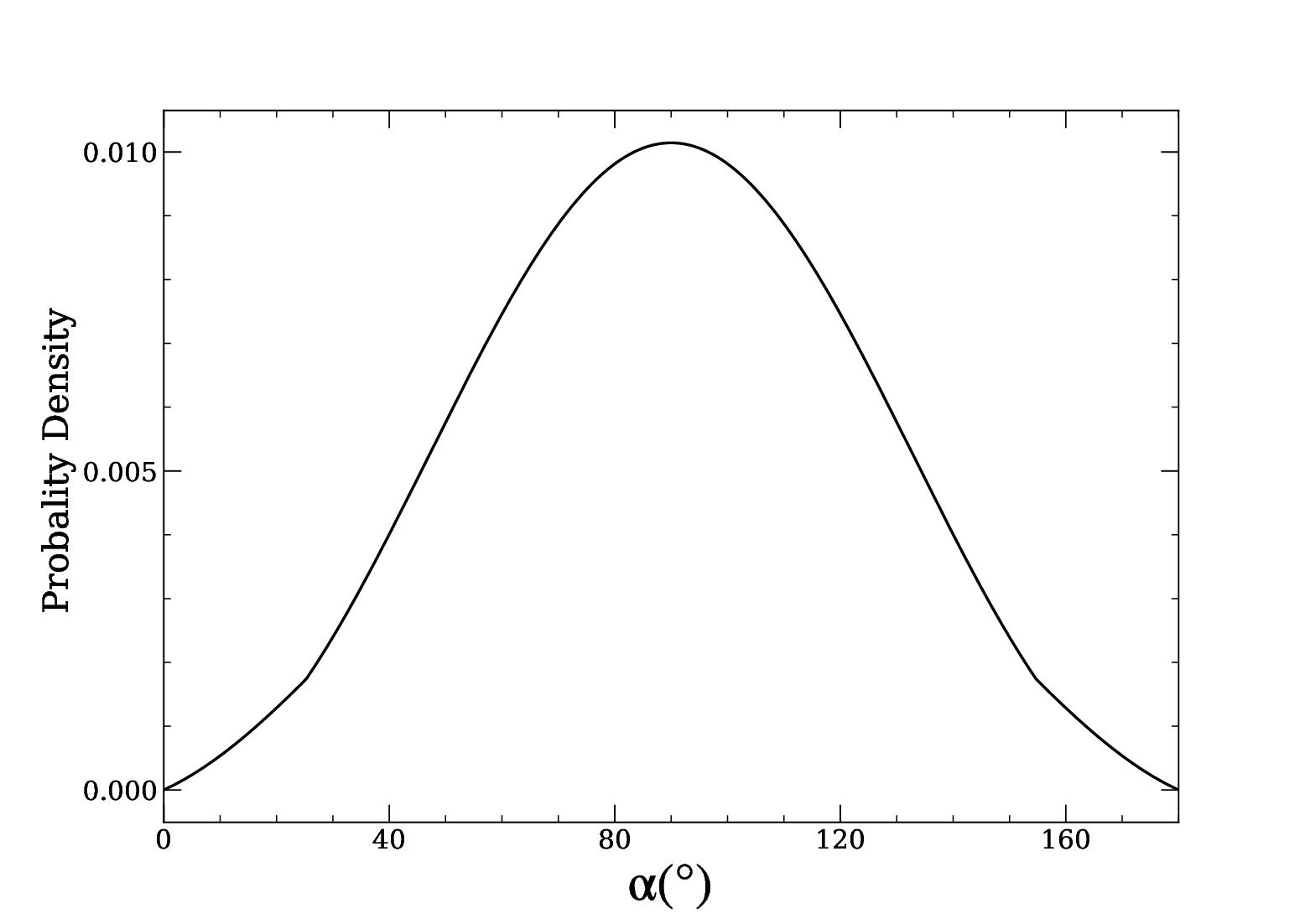}
    \caption{The probability density of magnetic inclination angle $\alpha$ derived from the period and pulse width of PSR B0950$+$08.}
    \label{alpha}
\end{figure}

\begin{figure}
    \centering
    \includegraphics[width = \columnwidth ]{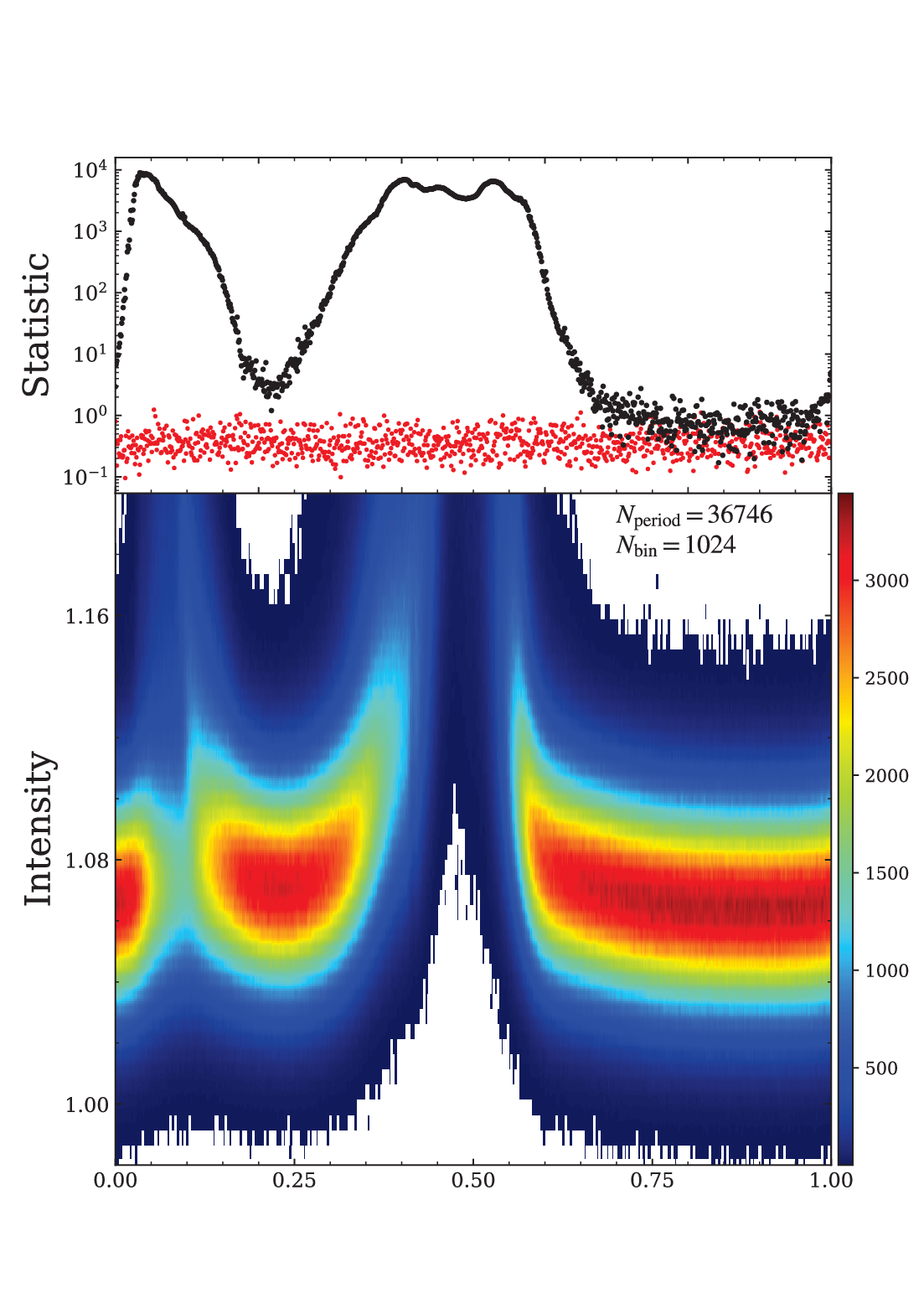}
    \caption{The radio signal distributions at each longitude are shown in the bottom panel, and the top panel shows the statistics of the Anderson-Darling test on them as the black points. The red points are the Anderson-Darling test statistics for the simulated random data with Gaussian distribution. And the number of single pulses with fixed radio emission intensity at each pulse phase is represented by different colors.}
    \label{stat}
\end{figure}
%
%\begin{figure}
%    \centering
%    \includegraphics[width = 1.\columnwidth ]{acft.eps}
%    \caption{Auto-correlation function of intensity fluctuations from the low-emission regions.}
%    \label{acf}
%\end{figure}
%
The $\Theta(n)$-function describes the statistical properties of the radio emission signal.
The region of the on-pulse (the pulse longitudes which have periodic and impulsive emission) and off-pulse can be distinguished by this function, because the value of the $\Theta(n)$-function is positive at on-pulse regions and goes to zero at the off-pulse regions (see Appendix~\ref{a}).
It should be noted that the fluctuation of the radio emission signals and noise may lead to the statistics error, and this error can possibly be eliminated if the number of the individual pulse ($N_{\mathrm{period}}$) is big enough.
This statistical properties implies that the value of the $\Theta(n)$-function may be influenced by
%the distribution of
the background noise (discussed in Appendix~\ref{b}).
A detailed behaviour of $\Theta(n)$-function at the entire pulse longitude is described as black curves in Fig.~\ref{fig7}, and the detailed property of the blue region in the middle panel is zoomed in the bottom panel.
From this figure, the result is obviously influenced by the background noise at the low-emission region.
To determine the emission properties of these regions, the values of $\Theta(n)$-function at the low-emission region need to be smoothed. The red curve in the bottom panel is the smoothed result via low-pass filtering, and apparently, this curve at the low-emission region is greater than zero (dash-dot gray reference line).
It is evident that, in the observed data, the values of the $\Theta(n)$-function at the two low-emission regions are positive.
The behaviour of this function strongly illustrates that radio emission signals fill the low-emission region, and this result supports the whole phase radiation picture that the emission signals fill the entire $360^{\circ}$ pulse longitude.
%Moreover, the values of this function at two concave structure regions are almost equal to each other, it demonstrates that the emission of the two minimum low levels have similar physical origin.
%
To better demonstrate that the $\Theta(n)$-function is equal to zero for continuous emission (the unpulsed emission) and is not affected by the non-linear instrumental response,
and assuming that the non-linear instrumental response may cause the radio signals of each pulse phase to follow gamma or
%
%log-normal distribution.We simulate random data with different distribution.
log-normal distribution, we simulate the $\Theta(n)$-function measurements for those distributions.
The behavior of $\Theta(n)$-function for these simulations are shown in Fig.~\ref{off} to compare with the properties of $\Theta(n)$-function for the bridge emission region of PSR B0950$+$08.
Panels (a), (b), (c) and (d) depict the gamma (with shape parameter of 3 and scale parameter of 1), gamma (with shape parameter of 5 and scale parameter of 1), log-normal (with shape parameter of 0.1 and scale parameter of 1) and Gaussian distribution, respectively, and the sample numbers of all simulated data are $10^5$.
As these figures show, the values of the $\Theta(n)$-function are consistent with zero for continuous emission and the non-linear instrumental response.
%%%%%%%%%%%%%%%%%%%%%%%%%%%%%%%%%%%%%%%%%%%%%%%%%%%%%%%%%%%%%%
\section{ conclusion and discussions}
\label{con}
In this work, the impulsive emission characteristics of PSR B0950$+$08 are detected for the first time over the whole pulse phase.
It is shown that the averaged pulse profile of this pulsar is consistent with previously presented works in the strong emission regions~\citep[e.g.][]{2004MNRAS.351..808M,2001ApJ...553..341E,2005MNRAS.364.1397J}.
From Figs.~\ref{fig1}, ~\ref{ppa} and ~\ref{fig7}, it is evident that the emission of the low-emission region (from the pulse phase 0.65 to 0.95, the emission properties at this pulse longitude are undetected and usually regarded as off-pulse regions in published literature) have been detected for PSR B0950$+$08.
With the detection of the emission at the bridge regions, the whole phase radiation characteristics of this pulsar are identified.
However, there are some issues which need to be discussed.

%\textbf{The inclination angle, $\alpha$, of the pulsar can be obtained by its polarization characteristics~\citep{ 1969ApL.....3..225R}. I n addition, assuming that the radio emission bean
%
%is symmetry with respect to the magnetic axis. The inclination angle can also be estimated by the pulse width \citep[e.g.][]{1984A&A...132..312G, 1988MNRAS.234..477L}
The radio emission of the PSR B0950$+$08 may be originated from the two poles, which is different from the other whole-phase radiation pulsar, PSR B0826$-$34.
%
%The whole phase radiation behaviour of this pulsar is different from that of PSR B0826$-$34.
%
\citet{2005MNRAS.356...59E} reported that the radiation extends through the whole pulse phase of PSR B0826$-$34 in the ``strong emission mode''.
The radio emission signals of PSR B0826$-$34 covers the whole phase could naturally be understandable because the inclination angle, $\alpha$, is extremely low ($\sim0.5^{\circ}$), as is evident from the large pulse width for this pulsar.
Whereas, the pulse width of strong emission for PSR B0950$+$08 is relatively small ($\sim31.4^\circ$), which implies a large inclination angle.
With the assumption of the radiation altitude as 1000\,km (at this altitude, the multi-pole field is weak and the field is dominated by the dipole component), one can calculate the maximum half width of the radiation beam, $\theta_{\mu}$ (corresponding to the radiation from the last open field lines), to be $\sim 25.8^\circ$.
Assuming that the directions of the magnetic axis and rotation axis are uniform and ignore the beam scale distribution, with the Eq. 2 in~\cite{2009ApJ...703..507L}, we estimate the probability density of magnetic inclination angle, $\alpha$, which is shown in Fig.~\ref{alpha}.
%
%It is obvious that the inclination angle can hardly be smaller than $10^\circ$ or larger than $170^\circ$.
It can be calculated that the probability density of the inclination angle (smaller that $10^\circ$ or larger than $170^\circ$) is $\sim 2.96 \times 10^{-3}$.
Moreover,
%the inclination angle measured through polarization observation of the strong emission is $\alpha \simeq 70^{\circ}$~\citep{2001ApJ...553..341E},
the inclination angle measured through polarization observation of the strong emission is $\alpha \simeq 75^{\circ}$~\citep{2001ApJ...553..341E}, which implies the radiation of PSR B0950$+$08 originates from both poles of an almost orthogonal rotator.
Therefore, the radio signals of the two bridge regions should be overlapped by radiation from the two poles.
%\textbf{Therefore, the results support that the radio emission of PSR B0950$+$08 come from the two poles.}

%\textbf{It is hardly to understand the origin radiation of the two low-emission regions by the radio coherent mechanism.}
The whole phase radiation implies that some radio emission from PSR B0950$+$08 must come from high altitude.
However, the assumption of radiation altitude at 1000\,km should be discussed carefully.
In fact, the mean pulse profile at lower frequency is much wider, with pulse width $\sim70^\circ$~\citep{2022A&A...658A.143B}, which leads to that the radiation altitude must be higher than 1500\,km under the assumption of $\alpha\simeq75^{\circ}$.
Furthermore, for the whole phase radiation, the maximum radiation altitude should be larger than 5000\,km even after the adjustment for the effects of aberration and retardation.
%
%Considering that the radius of the light cylinder is $\sim12000$\,km and that the inclination angle is $70^\circ$, we can derive an extremely high altitude of 5000\,km,
Considering the radius of the light cylinder for PSR B0950$+$08 of $\sim12000$\,km and its whole phase radiation characteristics, there must be a certain pulse longitude that the emission comes from the two poles at the same time, and it can be derived that the half width of at least one radiation beam, $\theta_{\mu}$, must not be less than $90^{\circ}$.
According to the radiation altitude equation (see Eq.(5) in ~\citet{2007A&A...465..525Z}), it is calculated that the emission height of a certain pulse longitude (probably in the two low-emission regions) is larger than $\sim 8000$\,km.
With the inclination angle is $75^\circ$, it can be also calculated, at least 40\% of the entire pulse period, the radiation comes from an extremely high altitude of 5000\,km,
%already reaching the location of ``outer gap'' to be popular for explaining the X-/$\gamma$-ray emission of pulsars~\citep{1986ApJ...300..500C}.
which is roughly the altitude of the outer gap where the high energy emission is thought to be originated~\citep{1986ApJ...300..500C}.
However, for the radio radiation of normal pulsars, the radiative particles are supposed to lose energy rapidly and not to keep radiating till reaching such a high altitude~\citep{1997ApJ...491..891Z}.
%
%For PSR B0950$+$08, the two bridges between main pulse and inter pulse most probably originate from this height, as a challenge
%to the origin and the coherence mechanism of the relativistic plasma for the radio emission.
%\textbf{Because at that high altitude, the coherent radio emission mechanism may be broken such as the coherent radio luminosity is not proportional to the square of the number of radiative particles or the energy of the particles can not support the radio emission.}
The radio emission is always regarded as being generated from low altitude of pulsars, due to two effects of the energy loss and the decoherence of the radiative particles which moving outwards along the opening magnetic field lines.
The radiation altitude larger that 8000\,km challenge the conventional particle acceleration and coherent radiation mechanism in pulsar magnetosphere discussed in~\citet{1975ApJ...196...51R}, which may imply additional acceleration electric field over the polar gap, i.e., the slot gap~\citep{1983ApJ...266..215A} or annular gap~\citep{2007ChJAA...7..496Q}.
%Moreover, the radio signals of the two bridge regions may be overlapped by radiation from the two-poles.

The whole pulse radiation characteristics is unveiled with the $\Theta(n)$-function, which exploits the variation property of the pulsar radiation.
In other words, the invariant radiation does not affect the analysis results.
For example, the stable emission from pulsar wind nebula which contributes off-pulse emission~\citep{2020MNRAS.495.2125R} is different from the whole phase radiation from pulsar.
In fact, the whole phase radiation is similar to the non-100\% pulsed fraction radiation of the high energy pulsar, in which the emission occurs over the whole phase while some specific phase regions are preferred.

The physical origin of the radio emission at different pulse longitudes is also noteworthy.
The averaged pulse profile shows complex shapes, and its radio emission beam is described by the so-called ``core-cone'' model.
According to this model, the radio radiation beam consists of a central core surrounded by two rings of nested conal emission, which are respectively named the inner and outer cones~\citep{1983ApJ...274..333R, 1990ApJ...352..247R}.
Although the whole pulse longitude exhibits radio emission, the emission properties vary with different pulse longitudes.
%The significance of this characteristics is exhibited in Figs.~\ref{ppa} and ~\ref{fig7}.
%
%The difference between inter-pulse and main pulse is described in Fig.~\ref{ppa}, and the well-resolved conal double structure of inter-pulse is different with the unresolved conal single structure of main pulse.
In Fig.~\ref{ppa} it can be seen that the well-resolved conal-double structure of the inter-pulse differs from the unresolved conal-single structure of main pulse.
In Fig.~\ref{fig7}, the values of $\Theta(n)$-function in the two low-emission regions are almost equal to each other, which may indicate similar emission physics.
%Fig.~\ref{fig7} reveals that the values of $\Theta(n)$ at two minimum low levels are almost equal to each other, and this result possibility indicates that similar emission properties appear at the two minimum low levels.
%Furthermore, the physical origin of the whole radiation characteristics is a critical question.

To determine the emission properties of the bridge regions, the distribution of the radio emission signals at certain pulse longitude is also considered.
This method reveal the emission properties through the distribution of the noise and radio emission at each pulse phase bin of the folded data, which follow different statistical laws.
Apparently, the fluctuation of the noise follows the Gaussian distribution, and this statistical property is different with the log-normal distribution of the radio emission signals of the normal pulse~\citep[][]{2001ApJ...563L..65C}.
%~\citep[][]{2004ApJ...610..948C}.
%
This property may cause that the distribution of the emission at the on-pulse phase interval is asymmetric.
The statistical results are shown in the bottom panel of Fig.~\ref{stat}, and the number of single pulses with fixed radio emission intensity at each pulse phase is represented by different colors.
From this figure, it faintly shows that the distribution of the upper and lower of the maximum distribution point (the red points) at each pulse phase bin is asymmetric.
%
%Moreover, the dispersion of the upper area is greater than the lower, and this distribution characteristics supports that there is radio emission components at the each pulse phase bin.
%
To better illustrate the asymmetry of the whole pulse longitudes, a longitude-resolved test statistics such as Anderson-Darling test is also used with the null hypothesis that the distribution of each longitude comes from Gaussian distribution~\citep[][]{Stephens1974}.
The Anderson-Darling statistics are shown in the top panel of Fig.~\ref{stat} as black points, and the red points represent the results, which are derived from the simulated random data with Gaussian distribution.
It demonstrates that the distribution at each longitude deviates the Gaussian distribution, which is consistent with the distribution characteristics in the bottom panel.
Although it seems that the above method can also reveal the emission in the bridge regions, the flux distribution at each phase bin may be affected by weak RFI in frequency domain and non-linear response of the instrument.
In contrast, the $\Theta(n)$-function would be better since that it just refers to the radiation variation characteristics in time domain.
%

%\textbf{Beside the method used above, the auto-correlation function (ACF) is widely used to reveal the structure characteristics of sub-pulse and micro-structure~\citep{1976ApJ...208..944C}. This method determines the emission characteristics via analyzing the emission correlation in a wide pulse longitude region (such as $\sim10 \%$ of the entire pulse period or the whole bridge region).
%For instance, the ACF of the first low-emission regions (from 0.17 to 0.30 pulse phase) or the second low-emission regions (from 0.65 to 0.95 pulse phase) can only unveil the time scale of the radio emission, can not detect the emission feature at each pulse longitude.}
%
%Thus the ACF can only determine the existence of the emission in the wide region instead of in every phase bin. In contrast, the $\Theta(n)$-function reveals the emission characteristics based on the emission correlation of any three adjacent phase bins, and can determine the radio emission at each pulse longitude.
Beside the method used above, the auto-correlation function (ACF) is widely used to reveal the structure characteristics of sub-pulse and micro-structure~\citep{1976ApJ...208..944C}.
In this method, it is needed to use several phase bins (the number of phase bin is usually larger than 10) in each single pulse to calculate the ACF result, and then the conclusion about the radiation existence on the phase interval of these bins can be drawn.
However, it is hard to distinguish if the radiation exists on a specific pulse phase due to the relatively long data length to calculate the ACF results.
In addition, the ACF may also be influenced by other correlations in the recorded data, i.e., the correlation introduced by the dedispersion process and the red component of the system noise.
In this paper, the $\Theta(n)$-function is introduced as method to judge the existence of the radiation at each phase bin.
To calculate $\Theta(n)$-function, the correlation of the intensity evolution (instead of the intensity) between adjacent phase bins is obtained to show the impulsive property of sub-pulse.
Therefore, the $\Theta(n)$-function represent the intensity correlation of the first order (while the classical ACF method reveal the zero order), which make it to avoid being influenced by the inaccuracy of the baseline.
In addition, the longitude-longitude cross correlation can also be used to determine the sub-pulse width~\citep{2001A&A...379..270K}.
However, this method is strongly influenced by the accuracy of the baseline in each single pulse and would lose efficacy for the weak emission of PSR B0950$+$08.

%\textbf{The $\Theta(n)$-function is just a method for detection of radio signals of weak emission for PSR B0950$+$08.}
In this work, the two low-emission regions of the averaged pulse profile for PSR B0950$+$08 exhibit significant concave structure rather that flat platform, which indicates that there is an impulsive radio emission at these two low-emission regions.
A $\Theta(n)$-function is proposed to quantitatively determine the radio signal of the weak emission regions, classifying an impulsive.
Through data analysis, the $\Theta(n)$-function is found to be a tentative method to determine the radio emission of the bridge regions for this pulsar, because it is just based on the emission correlation of any three adjacent phase bins.
However, it would be noted that given the time-scale discussed in Appendix~\ref{a}, the $\Theta(n)$-function
%not an universal determination radio emission method.
%\textbf{not the only method for detection of radio emission}.
may not be suitable for detecting the radiation of all pulsars.
%since the time-scale of the fine structure discussed in Appendix~\ref{a}.
It should be caution that if one applies the $\Theta(n)$-function to determine the radio signal in the weak emission regions for other pulsars.

It is difficult to determine the baseline position of PSR B0950$+$08, because the baseline determination for this pulsar is affected by many problems.
For instance, the conventional baseline subtraction is not suitable for this pulsar.
Secondly, the change of the baseline with time will affect its determination.
%but this effect can be eliminated through the flattening of the baseline evolution in each individual pulse with considering the narrow pulse phase property of the individual pulse.
%
Moreover, the radiation intensity of this pulsar is influenced by the effect of the interstellar scintillation~\citep[e.g.][]{2022A&A...658A.143B},
%\citep[][]{2008ARep...52..736S,1985ApJ...288..221C},
which could also influence determining process.
In fact, the baseline emission is much more difficult to remove completely for the single-dish radio telescope.
In this work,
%the -function is a tentative method to detect the impulsive radio signal of the ¡°bridge¡± emission region of PSR B0950+08.
%
the detection result of the $\Theta(n)$-function is hardly affected by the baseline emission of PSR B0950$+$08 for this observation,
%effect of the inaccuracy of the baseline,
since it only depends on the pulsed radio signal of the sub-pulse of this pulsar. The baseline emission of the pulsar can be determined by the radio interferometer through analysing the spatial variation of the background emission.
More precise measurements for the baseline emission of this pulsar may be provided by the future interferometers such as the SKA~\citep{2015aska.confE.171C} or the FAST extension array (FASTA).

The determination of the accuracy baseline is very important for the pulsar with the detectable emission over the whole pulse phase, since the PPA is related to the radiation geometry.
But the PPA of the whole phase radiation is hard to be calibrated because it is sensitive to the baseline determination in the weak radiation phase region.
The polarization features of this pulsar are not presented in this paper, since we have not made satisfied calibration during the observations.
The non-linear evolution of baseline response with time severely influenced the calibration.
Therefore, real time calibration information is needed to determine the baselines of the Stokes parameters precisely, and the optimized calibration of this pulsar has already been planned.
%
%To determinate the baseline position of the pulsar with detectable emission over the whole pulse phase, one potential baseline determination method is that considering the relationship between the baseline and single pulse.
%
%The baseline changing with time will affect the baseline determination, but this effect can be eliminated through the flattening of the baseline evolution in each individual pulse with considering the narrow pulse phase property of the individual pulse.}
%

%One possible concern is
The effects of saturation from the digitization process, as presented in~\citet{1998PASP..110.1467J}.
There would be an artifact caused by the analog-to-digital conversion and dispersion, which may bring a scattered power into the data.
This effect will result in a fake correlation along pulse phase and a wider radiation phase, and the range of this effect depends on the dispersion.
%
%the first artifact is introduced by the non-linearities that appear during analog-to-digital conversion, and another artifact is caused by the ``quantization noise'',
It could be calculated that the dispersion sweep in our frequency range of 1050-1450 MHz is 5.3\,ms, $\sim 2 \%$ of the entire pulse period for PSR B0950$+$08.
Thus, with the digitization effect, the radio pulse would be widened by 2\% and an artificial correlation within 2\% pulse period would be produced.
The non-zero value of the $\Theta(n)$-function may be also induced by the time correlation.
We expect that future interferometers such as the SKA or the FASTA will verify the current findings.
%
%pointed that there exist mainly two artifacts during the digitization process, the first artifact is introduced by the non-linearities that appear during analog-to-digital conversion, and another artifact is caused by the ``quantization noise'', it would bring a scattered power into the data. However, all of observation data have been checked with caution via time-frequency dynamics spectrum, the whole pulse longitudes, especially the weak emission regions are not distorted. And, the radio signal of the weak emission region is possible little affected by the effect of dispersive delay, but it is only $2 \%$ of the entire pulse period for PSR B0950$+$08. Therefore, the whole pulse phase radiation characteristics is detected over at least 98$\%$ of the rotation period in this work.}
%
%The whole phase radiation phenomena of PSR B0950$+$08 is detected by the FAST with high sensitivity, and in the future, the Square Kilometre Array (SKA) \citep{2015aska.confE.171C} will also support the excellent observational window to provide more strong ability to make a depth and detailed studies for radio pulsars.
%
Additionally, we anticipate that FAST and these future interferometers are potentially detecting more pulsars with whole phase radiation in the future.
%
%Furthermore, in view of the fact that the polarization behaviors of repeating fast radio bursts have been investigated extensively with FAST too~\citep{WangWY2022,WangWY2021}, a new era of studding pulsar electrodynamics and related radiative mechanisms is expected.
%
%The Acknowledgements section is not numbered. Here you can thank helpful
%colleagues, acknowledge funding agencies, telescopes and facilities used etc.
\section*{acknowledgements}
This work made use of the data from FAST (Five-hundred-meter Aperture Spherical radio Telescope). FAST is a Chinese national mega-science facility, operated by National Astronomical Observatories, Chinese Academy of Sciences.
This work is supported by the National Natural Science Foundation of China (Grant No. 12003047, 12133003), the National SKA Program of China (No. 2020SKA0120100) and the strategic Priority Research Program of CAS (XDB23010200).

%%%%%%%%%%%%%%%%%%%%%%%%%%%%%%%%%%%%%%%%%%%%%%%%%%
\section*{Data Availability}
 The data underlying this work are available in the FAST project PT2020$-$0034, and can be shared on request to the FAST Data Center.


\bibliographystyle{mnras}
\bibliography{ref} % if your bibtex file is called example.bib

%Alternatively you could enter them by hand, like this:
%This method is tedious and prone to error if you have lots of references
%\begin{thebibliography}{99}
%\end{thebibliography}

%%%%%%%%%%%%%%%%%%%%%%%%%%%%%%%%%%%%%%%%%%%%%%%%%%

%%%%%%%%%%%%%%%%% APPENDICES %%%%%%%%%%%%%%%%%%%%%

\appendix
\section{the property of the $\Theta$-function}
\label{a}
Detailed expression of the $\Theta(n)$-function is shown as equation(\ref{eq1}), and here we will present that the values of $\Theta(n)$-function at the emission regions are positive.

For the data, at any pulse phase interval with emission, the intensity of the noise follows the Gaussian distribution.
%
%The statistics properties of the radio emission signals and noise is described by the function $\Theta(n)$, this properties ensures that the behaviour of the function $\Theta(n)$ is not influenced by the distribution of the noise intensity at the each pulse phase. So these turns which are contributed by the noise intensity can be ignored, when analyzed the properties of the function $\Theta(n)$.
%
The properties of $\Theta(n)$-function can be described by the follows,
\begin{equation}
    \begin{split}
        & \Theta(n) = \sum_{k=1}^{N_{\mathrm{period}}} \bigg\{2 \left( I_{k,n-1} - I_{k,n+1}  \right)^2 \\
        & - \left [ \left( I_{k,n-1} - I_{k,n} \right)^2  + \left( I_{k,n} - I_{k,n+1} \right)^2 \right] \bigg\}\\
        & =  \sum_{k =1}^{N_{\mathrm{period}}} \bigg\{2 \left( S_{k,n-1} - S_{k,n+1} + \sigma_{k,n-1} - \sigma_{k,n+1} \right)^2 \\
        & - \left[ \left( S_{k,n-1} - S_{k,n} + \sigma_{k,n-1} - \sigma_{k,n} \right)^2  \right] \\
        & - \left[ \left( S_{k,n+1} - S_{k,n} + \sigma_{k,n+1} - \sigma_{k,n} \right)^2 \right]\bigg\} \\
%        & = \sum_{k =1}^{N_{\mathrm{period}}} \bigg\{ 2\left(S_{k,n-1} - S_{k,n+1} \right)^2 - [ \left(S_{k,n-1} - S_{k,n} \right)^2 + \left( S_{k,n} - S_{k,n+1} \right)^2 ] \bigg\}  \\
%        & + \sum_{k =1}^{N_{\mathrm{period}}} \bigg\{  2\left( \sigma_{ k,n-1} - \sigma_{ k,n+1} \right)^2  \\
%        & - \left[ \left( \sigma_{ k,n} - \sigma_{ k,n-1} \right)^2 + \left( \sigma_{ k,n} - \sigma_{ k,n+1}\right)^2 \right] \bigg\} \\
%        & + \sum_{k =1}^{N_{\mathrm{period}}} 2 \bigg\{ 2 \left( S_{k,n-1} - S_{k,n+1} \right) \left( \sigma_{k,n-1} - \sigma_{k,n+1} \right) \\
%        & - \left( S_{k,n-1} - S_{k,n} \right) \left( \sigma_{k,n-1} - \sigma_{k,n} \right) \\
%        & - \left( S_{k,n+1} - S_{k,n} \right) \left( \sigma_{k,n+1} - \sigma_{k,n} \right) \bigg\} \\
        & \approx \sum_{k =1}^{N_{\mathrm{period}}} \bigg\{ 2\left(S_{k,n-1} - S_{k,n+1} \right)^2 - \\
        & \left[ \left(S_{k,n-1} - S_{k,n} \right)^2 + \left( S_{k,n} - S_{k,n+1} \right)^2 \right] \bigg\},
    \end{split}
\end{equation}
where $S_{k,n}$ and $\sigma_{k,n}$ are the emission signal and noise intensity contributed by the $k$-th pulse and $n$-th pulse phase bin,respectively.

For the observation data, a same structure of the sub-pulse repeat many times in a large number of cycles, so that a pulse phase is ergodic in all positions of sub-pulse.
In other words, the summation of all cycles is equivalent to that of all points in sub-pulses with different shape,
\begin{equation}
    \begin{split}
        &\Theta(n) \approx \sum_{k =1}^{N_{\mathrm{period}}} \frac{1}{L} \sum_{l=1}^{L} \bigg\{ 2 \left( S_{k,n,l-1} - S_{k,n,l+1} \right)^2 \\
        & - \left[ \left( S_{k,n,l} - S_{k,n,l-1}\right)^2 + \left( S_{k,n,l} - S_{k,n,l+1} \right)^2 \right] \bigg\}\\
        & \approx \sum_{k =1}^{N_{\mathrm{period}}} \frac{1}{L} \sum_{l=1}^{L} \bigg\{ \left( S_{k,n,l} - S_{k,n,l-1} \right)^2 + \left(S_{k,n,l+1} - S_{k,n,l} \right)^2 \\
        & + 4\left(S_{k,n,l} - S_{k,n,l-1} \right) \left(S_{k,n,l+1} - S_{k,n,l} \right) \bigg\} \\
        & \approx \sum_{k =1}^{N_{\mathrm{period}}} \frac{1}{L} \sum_{l=1}^{L} \left(\Delta_{l}^2 + \Delta^2_{l-1} + 4 \Delta_{l}\Delta_{l-1} \right)
    \end{split}
\end{equation}
where $S_{k,n,l}$ is the $l$-th bin of the sub-pulse with $L$ bins of the $k$-th pulse and $n$-th pulse phase bin,
%\begin{equation}
%    \begin{split}
%        & \Theta(n) = \sum_{k =1}^{N_{\mathrm{period}}} \frac{1}{L} \sum_{l=1}^{L} \bigg\{ \left( S_{k,n,l-1} - S_{k,n,l+1} \right)^2 \\
%        & + 2\left(S_{k,n,l} - S_{k,n,l-1}\right) \left(S_{k,n,l+1} - S_{k,n,l}\right) \bigg\} \\
 %       & = \sum_{k =1}^{N_{\mathrm{period}}} \frac{1}{L} \sum_{l=1}^{L} \bigg\{ \left( S_{k,n,l} - S_{k,n,l-1} \right)^2 + \left(S_{k,n,l+1} - S_{k,n,l} \right)^2 \\
%        & + 4\left(S_{k,n,l} - S_{k,n,l-1} \right) \left(S_{k,n,l+1} - S_{k,n,l} \right) \bigg\} \\
%        & = \sum_{k =1}^{N_{\mathrm{period}}} \frac{1}{L} \sum_{l=1}^{L} \left(\Delta_{l}^2 + \Delta^2_{l-1} + 4 \Delta_{l}\Delta_{l-1} \right)
%    \end{split}
%\end{equation}
$\Delta_{l} = S_{k,n,l+1} - S_{k,n,l}$ is the first order difference of the sub-pulse.

\begin{figure}
    \centering
    \includegraphics[width = 1.\columnwidth ]{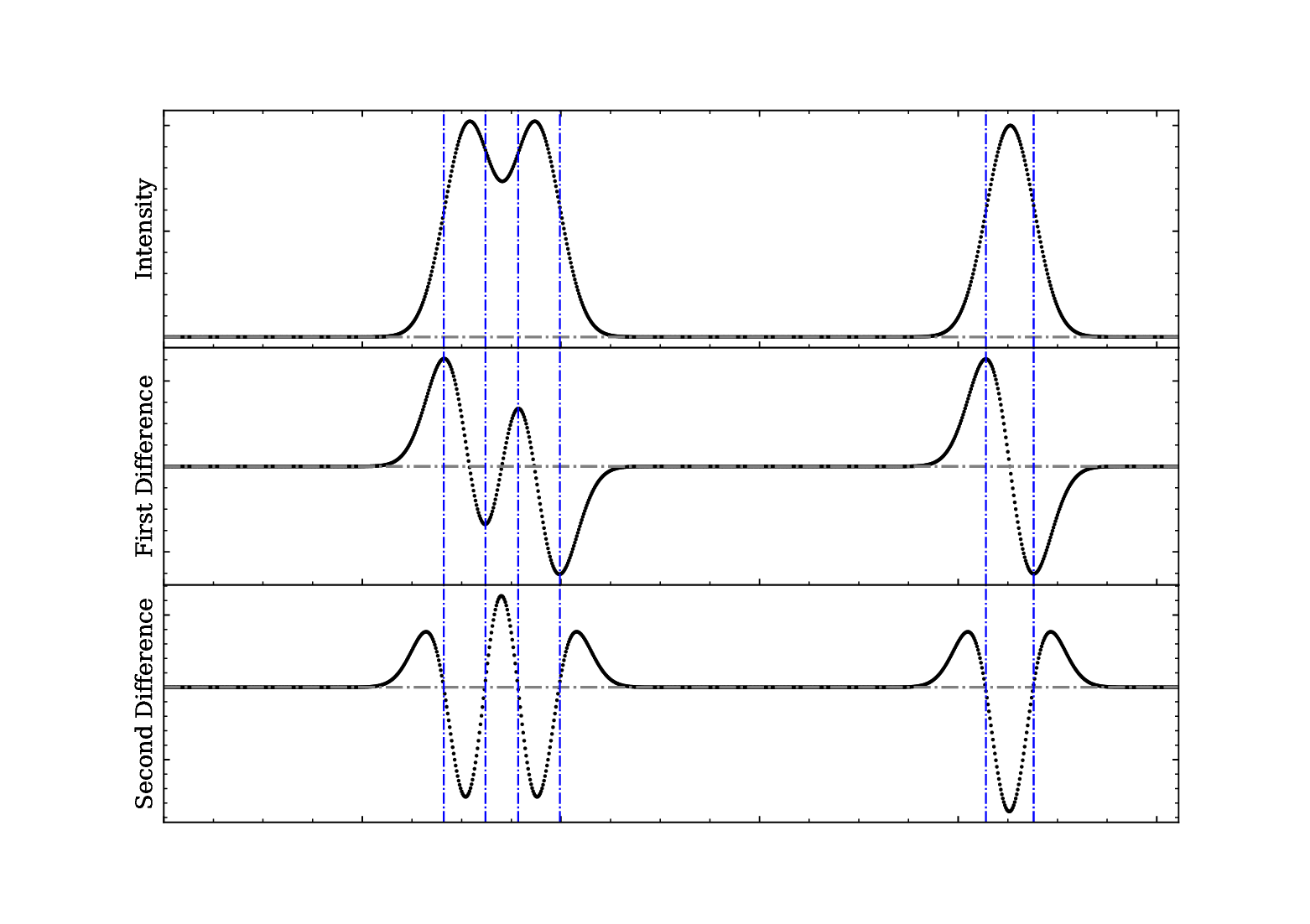}
    \caption{Intensity profile shape and its differences, and the dot-dashed gray line represents the reference zero line. The vertical dot-dashed lines divide the profile into several segments based on the sign of the second order difference.}
    \label{p0m}
\end{figure}

In this work, the time resolution of the folded data is $\sim$0.25\,ms.
According to the previous work, the characteristics time-scale of the fine structure is $\sim$1\,ms for PSR B0950$+$08~\citep[][]{2016MNRAS.455..150U}.
In other words, the second order difference can hardly change sign in succession in 1\,ms ($\sim$4 phase bins) for the folded data.
In the interval that the second order difference has same sign (the bin index of the start and end of the interval are $l_{\mathrm{s}}$ and $l_{\mathrm{e}}$,respectively), $\Delta_l$ is monotonic.
Fig.~\ref{p0m} shows an example of the profile shape and its differences at first and second order in the top, middle and bottom panels, respectively, and the vertical dot-dashed lines in the figure divide the profile into several segments.
It can be seen that second order difference in each segment has same sign, and the first order difference is monotonic.
%It can be seen that the same sign interval of second order difference can not affected by the intensity profile shape whether it is a well-resolved double peak structure or a single peak structure.
%
If the sign of $\Delta_{l}$ does not change,
we have $ M = \displaystyle \sum_{l=l_{\mathrm{s}}}^{l_{\mathrm{e}}} \left(\Delta_{l}^2 + \Delta^2_{l-1} + 4 \Delta_{l}\Delta_{l-1} \right) \ge 0 $. Otherwise, if $\Delta_{l}$ changes signs, without loss of generality, suppose $\Delta_{l_{\mathrm{s}}} \ge ... \ge \Delta_{m} > 0 \ge \Delta_{m+1} \ge ... \ge \Delta_{l_{\mathrm{e}}}$, it follows that
\begin{equation}
    \begin{split}
      & \sum_{l_{\mathrm{s}}}^{l_{\mathrm{e}}} \left(\Delta_{l}^2 + \Delta^2_{l-1} + 4 \Delta_{l}\Delta_{l-1} \right) \\
      & = \sum_{l = m-1}^{l_{\mathrm{s}}} \left(\Delta_{l}^2 + \Delta^2_{l-1} + 4 \Delta_{l}\Delta_{l-1} \right) + \left(\Delta_{m}^2 + \Delta^2_{m-1} + 4 \Delta_{m}\Delta_{m-1} \right) \\
      & + \left(\Delta_{m+1}^2 + \Delta^2_{m} + 4 \Delta_{m+1}\Delta_{m} \right) + \sum_{l = m+2}^{l_{\mathrm{e}}} \left(\Delta_{l}^2 + \Delta^2_{l-1} + 4 \Delta_{l}\Delta_{l-1} \right) \\
      & \ge \left(\Delta_{m}^2 + \Delta^2_{m-1} + 4 \Delta_{m}\Delta_{m-1} \right) + \left(\Delta_{m+1}^2 + \Delta^2_{m} + 4 \Delta_{m+1}\Delta_{m} \right) \\
      & = \Delta_{m-1}^2 - 2 \Delta_m^2 + 4 \Delta_m\Delta_{m-1} + (\Delta_{m+1} + 2\Delta_{m})^2 \\
      & \ge \Delta_{m-1}^2 - 2 \Delta_m^2 + 4 \Delta_m\Delta_{m-1} \\
      & \ge \Delta_{m}^2 - 2 \Delta_{m}^2 + 4 \Delta_{m}^2 \\
      & = 3 \Delta_{m}^2 >0.
    \end{split}
\end{equation}
To sum up, $\Theta(n) > 0$ at the interval of the emission pulse phase.

\section{The effects of the background noise}
\label{b}

\begin{figure}
    \centering
    \includegraphics[width = 1.\columnwidth ]{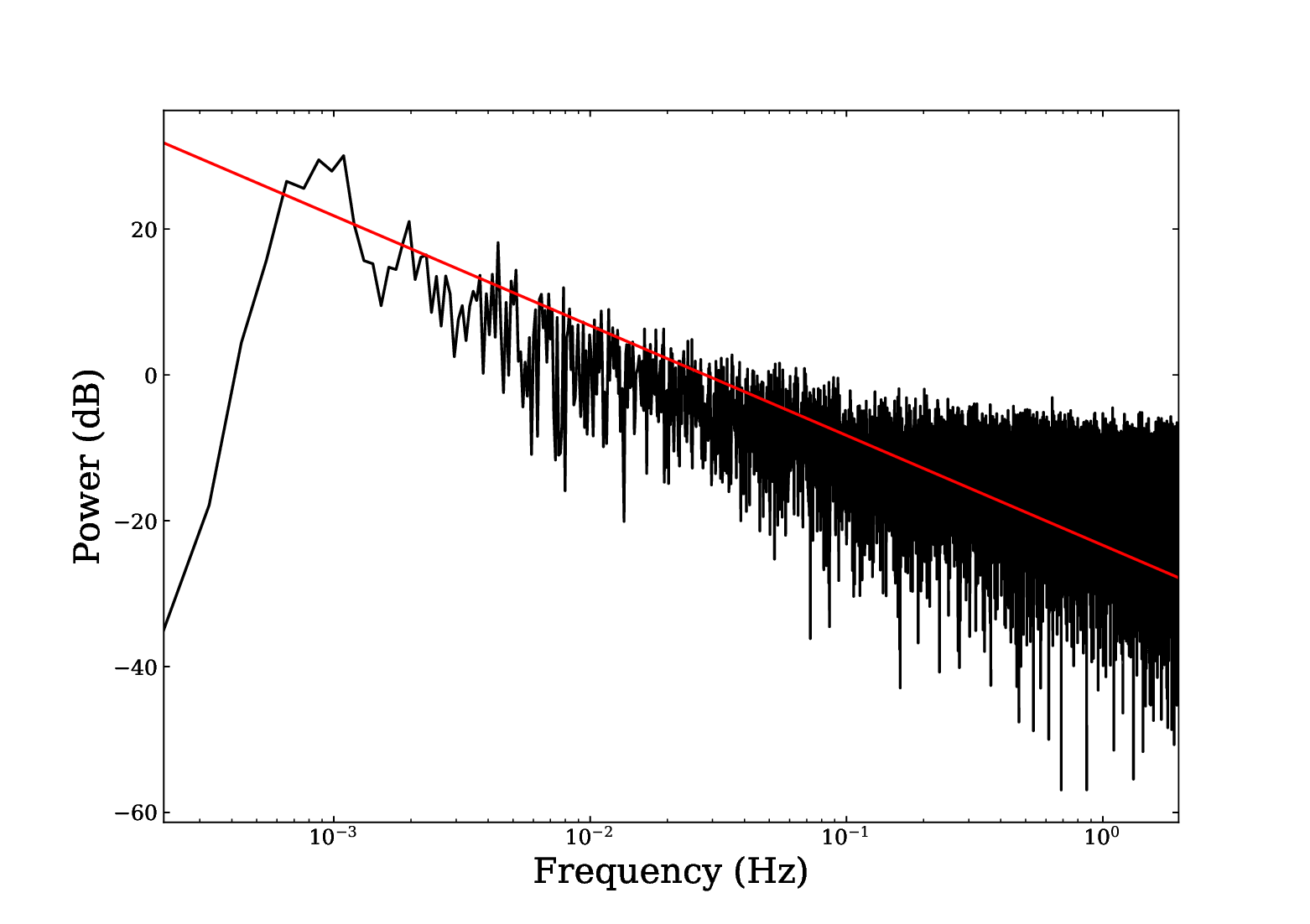}
    \caption{The spectrum of the background noise intensity is normalized by its white noise intensity, and the red line corresponds to the power-law fitting. The background noise intensity is obtained by averaging the minimum 10 phase bins in each period.}
    \label{spec}
\end{figure}

\begin{figure}
    \centering
    \includegraphics[width = 1.\columnwidth ]{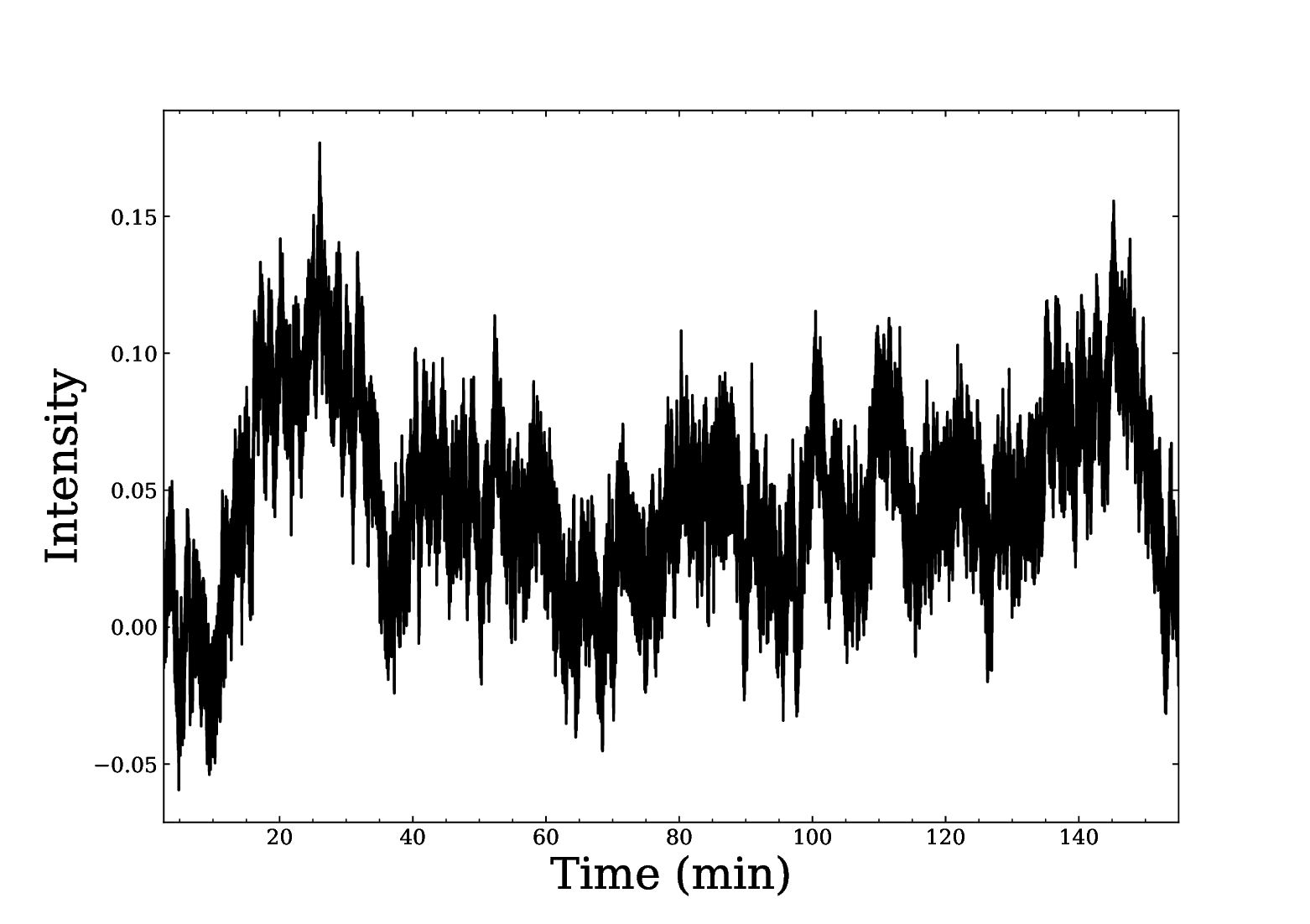}
    \caption{Simulated noise according to the power-law of $P(f) = 3.969 \times 10^{-3} f^{-1.506}$, and it has same the white noise intensity as the background noise.}
    \label{N}
\end{figure}

\begin{figure}
    \centering
    \includegraphics[width = 1.\columnwidth ]{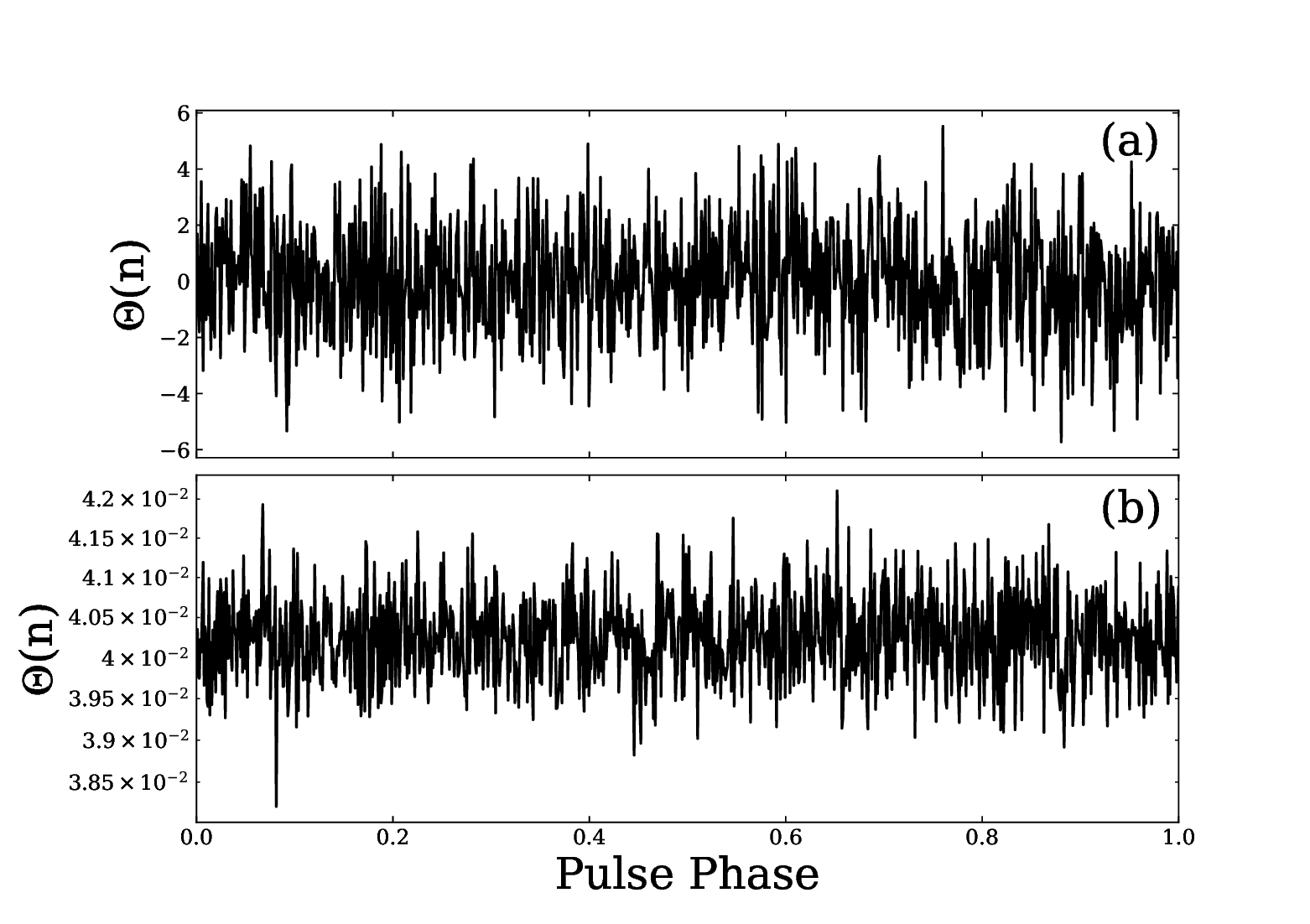}
    \caption{The $\Theta(n)$-function of the simulated background noise and its red noise components are shown in panels (a) and (b), respectively.}
    \label{D}
\end{figure}

When using the $\Theta(n)$-function to quantitatively determine the emission characteristics of the weak regions, such as the two low-emission regions of PSR B0950$+$08 in this work, one concern is the effects of the background noise. particularly the red noise components.
Fig.~\ref{spec} shows the spectrum of the background noise intensity relative to the white noise, and red line corresponds to the power-law fitting of the red noise component, $P(f) = 3.969 \times 10^{-3} f^{-1.506}$.
With the power-law expression, the background noise is simulated, which has same white noise intensity as the observed data.
The simulated red noise is shown in Fig.~\ref{N}.
The values of the $\Theta(n)$-function of the simulated noise is calculated, which is presented in the panel (a) of the Fig.~\ref{D}.
In addition, for comparison, the $\Theta(n)$-function of the pure red noise component is shown in the panel (b).
It can be seen that the positive property of the $\Theta(n)$-function is hardly affected by the red noise components in this observation, and the fluctuations of the $\Theta(n)$-function in Fig.~\ref{fig7} are almost resulted from the background noise.

%\textbf{To take into account the effect of the dispersion measure (DM) for this pulsar, the properties of the the $\Theta(n)$-function at the relatively minimum level varies with the sampling time of the fold data are shown in Fig.~\ref{B4}.
%
%It is shown that the values of the $\Theta(n)$-function are hardly affected by the timescale of the effects of the dispersion measure ($\approx 5.0$ or $10.0$ \,$\mu s$).
%
%Moreover, the none zero value of the $\Theta(n)$-function maybe also introduced by short timescale correlation.
%
%We expect that future high sensitivity interferometry such as SKA or FAST extension array (FASTA) will verify the current findings.
%}
%show, it can be seen that the properties of the $\Theta(n)$-function are not affected by the red noise components in this observation.

%
%\begin{figure}
%    \centering
%    \includegraphics[width = 1.\columnwidth ]{V0.eps}
%    \caption{The $\Theta(n)$-function of the relatively minimum level varies with the sampling time of the fold data.}
%    \label{B4}
%\end{figure}
%\textbf{To exhibit the micro-structure emission characteristics of PSR B0950$+$08, Fig.~\ref{p0m} shows that the complex sub-pulse structure and its first derivation of a certain period, and the properties of the $\Theta(n)$ function depends on the sub-pulse structure of all detected period, but the interval of the fine structure of this pulsar is surely depicted, because the effect of noise fluctuation is still affecting these structures.}
%t can be placed in an Appendix which appears after the list of references.

%%%%%%%%%%%%%%%%%%%%%%%%%%%%%%%%%%%%%%%%%%%%%%%%%%

% Don't change these lines
\bsp	% typesetting comment
\label{lastpage}
\end{document}